\documentclass[conference]{IEEEtran}

\usepackage{graphicx}
\usepackage{subfig}
\usepackage{booktabs}
\usepackage{subfig}
\usepackage{listings}
\usepackage{color}
\usepackage{multirow}
\usepackage{url}
\usepackage{array}
\usepackage{ctable}
\usepackage{amsmath}
\newtheorem{definition}{Definition}

\newcommand{\code}[1]{{\small\textsf{#1}}}
\newcommand{\groum}{GROUM}

\newcommand{\one}{XRank}
\newcommand{\two}{XHand}

\newcommand{\toolex}{ExAssist}
\newcommand{\tool}{ExAssist}
\newcommand{\firstmodel}{XRank}
\newcommand{\secondmodel}{XHand}

\author{\IEEEauthorblockN{Tam The Nguyen,
Phong Minh Vu,
Tung Thanh Nguyen}

\IEEEauthorblockA{Department of Computer Science and Software Engineering\\
Auburn University\\
\{tam,lenniel,tung\}@auburn.edu}\\
}

\begin{document}
\title{Recommendation of Exception Handling Code in Mobile App Development}
\maketitle

% abstract

\begin{abstract}
	In modern programming languages, exception handling is an effective mechanism to avoid unexpected runtime errors. Thus, failing to catch and handle exceptions could lead to serious issues like system crashing, resource leaking, or negative end-user experiences. However, writing correct exception handling code is often challenging in mobile app development due to the fast-changing nature of API libraries for mobile apps and the insufficiency of their documentation and source code examples. Our prior study shows that in practice mobile app developers cause many exception-related bugs and still use bad exception handling practices (e.g. catch an exception and do nothing). To address such problems, in this paper, we introduce two novel techniques for recommending correct exception handling code. One technique, XRank, recommends code to catch an exception likely occurring in a code snippet. The other, XHand, recommends correction code for such an occurring exception. We have developed {\tool}, a code recommendation tool for exception handling using XRank and XHand. The empirical evaluation shows that our techniques are highly effective. For example, XRank has top-1 accuracy of 70\% and top-3 accuracy of 87\%. XHand's results are 89\% and 96\%, respectively.
\end{abstract}

\section{Introduction}
Exceptions are unexpected errors occurring while a software system is running. For example, when a software system wants to open a file with a given file name but, unexpectedly to the programmer, the file system contains no file having that name, a \emph{"file not found"} exception occurs. Failing to handle exceptions properly could lead to more serious errors and issues such as system crashes or resource leaks. For example, a prior study reports that correctly releasing resources in the presence of exceptions could improve 17\% in performance of the application~\cite{weimer_mistakes}. Thus, effective exception handling is important in software development.

Modern programming languages like Java or C++ often provide built-in support for exception handling. For example, in Java, we can wrap a \code{try} block around a code fragment where one or more exceptions potentially occur. Then, we add one or more \code{catch} blocks to handle such exceptions. An API library often defines many API-specific exception types and exception handling rules. For example, in Java SDK, class \code{FileNotFoundException} is defined for the \emph{file not found} exception. When such an exception happens, the software system could notify users about the error and/or write the relevant information (e.g., filename) to the system's log for future debugging or investigations. 

However, learning to handle exception properly is often challenging especially in mobile app development due to several reasons. First, a mobile API library often consists of a large number of components. For example, the Android application framework contains over 3,400 classes, 35,000 methods, and more than 260 exception types \cite{Linares_api}. Moreover, the documentation for handling exceptions is generally insufficient. Kechagia \textit{et al.}~\cite{Kechagia_undocumented} found that 69\% of the methods from Android packages in their stack traces had undocumented exceptions in the Android API and 19\% of the crashes could have been caused by insufficient documentation. Coelho \textit{et al.}~\cite{hazard} found that documentation for explicitly thrown \code{RuntimeExceptions} is almost never provided. Second, due to the fast development of mobile devices and the strong competition between software and hardware vendors, those frameworks are often upgraded quickly and include very large changes and therefore they could introduce new types of exceptions. For example, 17 major versions of Android framework containing nearly 100,000 method-level changes have been released within five years~\cite{Linares_api}. Thus, it is difficult to learn and memorize what method could cause what exception and what to do when a particular exception occurs. That could lead to a high number of programming errors (bugs) related to exceptions in app development.

In our previous work~\cite{currentwork}, we performed a detailed empirical study on 246 exception-related bugs and fixes from 10 Android apps. We discovered several interesting findings on the nature of exception-related bugs in mobile app development and how app developers fix them. For example, we found out that almost all exception bugs (80\%) cause serious problems for the apps such as crashing or running in an unstable state. Furthermore, our result shows that programmers do not perform any actions to handle occurring exceptions in about 16\% bug fixes and they add repairing code in only 42\% of the exception bug fixes. In addition, app developers even failed to handle exceptions properly in several exception bug fixes. These findings suggest that forgetting to catch an exception or catching it and doing nothing are bad practices and could cause severe consequences (e.g. system crashes). Therefore, an automated tool that can recommend catching and handling exceptions properly would be useful for app developers.

In this paper, we propose two techniques to recommend proper exception handling. {\one} alerts us to catch exceptions for a given code fragment and recommends a ranked list of exception types to catch. {\two} recommends what methods to call in the \code{catch} block if we want to repair/reset the object' states when the exception happens. 

{\one} models associations between exceptions and method calls in training data based on fuzzy set theory. Once trained, given a code fragment (which would appear in the \code{try} block), {\one} extracts its set of method calls and recommends a rank list of exceptions that could appear based on those fuzzy sets.

{\two} is a statistical language model for learning and recommending exception handling method call sequences. {\two} simplifies the problem of modeling exception handling method sequences as the natural language problem of predict missing words on a sentence. With this assumption, {\two} makes use of $n$-gram model and multi-class classification model to train and predict exception handling methods. The details of our techniques are presented in Section III. 

Based on our proposed techniques, we developed {\tool}, a code recommendation tool for exception handling. {\tool} predicts what types of exception could occur in a given piece of code and recommends proper exception handling code for such exceptions. When requested, it will add such code into the current code editor. {\tool} is released as a plugin of IntelliJ IDEA and Android Studio, two popular IDEs for Java programs and Android mobile apps. The details of the tool are presented in Section IV.

We have conducted several experiments to evaluate the usefulness and effectiveness of our recommendation techniques on a dataset containing over 10 million methods from over 4 thousands mobile apps downloaded from Google's Android App store. The evaluation results of our techniques are promising. For example, {\one} has the top-1 accuracy of 70\% and top-5 accuracy of 93\% when recommending exceptions to catch. {\two} can recommend repairing method calls with top-1 accuracy of 89\% and top-3 accuracy of 96\%. We also evaluated our exception handling tool, {\tool}, on the tasks of detecting and fixing real exception handling related bugs. Over 128 exception related bugs, {\tool} correctly detects occurrences of exceptions on 90\% of the cases. It recommends the exception types in top-1 recommendation with 75\% accuracy and top-3 accuracy of 90\%. In exception handling code of 42 exception related bug fixes, the recommendations by {\tool} match the fixes of developers in 64\% of the cases. In addition, {\tool} vastly outperforms the baseline on the recommendation task with the improvement of 40\%. The details of our evaluation process are presented in Sections V. 

The key contributions of our paper include:
\begin{itemize}
	\item A technique for recommending potential exceptions to catch in a given a code fragment,
	\item A technique for recommending method calls to repair/reset object states when an exception occurs,
	\item A code recommendation tool for exception handling for Android Studio,
	\item An extensive evaluation of those recommendation techniques and the tool on a large dataset.
\end{itemize}

The remaining of this paper is organized as follows. Section II summarizes our motivational study on bugs and fixes related to exceptions. Section III presents our two exception-handling recommendation techniques, details of the training and recommendation process, and the design overview of {\tool}. Section IV introduces the usages of our exception handling recommendation tool, {\tool}. We present our evaluation in Section V and discuss the related works in Section VI.

\section{Motivation}
\subsection{Exception bugs and fixes}
\begin{figure}[h]
	\centering
	\includegraphics[scale = 0.45]{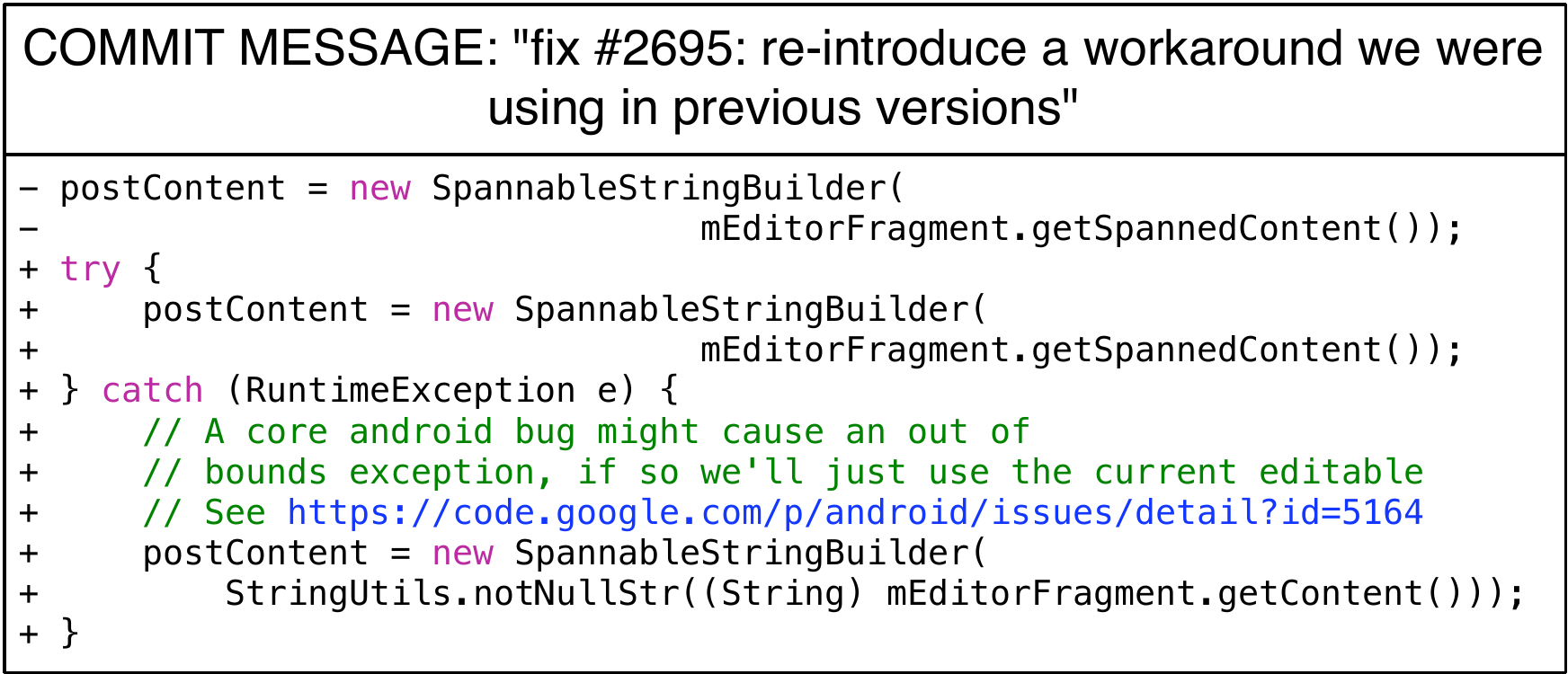}
	\caption{An exception bug fix appears in the \code{Wordpress} project}
	\label{empirical:example1}
\end{figure}

In this section, we present our motivation for solving the problem of exception handling for mobile app development. We focus on studying bugs that are caused by not catching exceptions or adding proper exception handlers. For convenience, we defined those bugs as \textbf{exception bugs}. The fix for those type of bugs is called \textbf{exception bug fixes}. Figure \ref{empirical:example1} shows a change of the an exception bug fix appears in the \code{Wordpress} project\footnote{http://android.wordpress.org}. In the example, the bug happened because of an unhandled \code{RuntimeException} occurs while creating a new \code{SpannableStringBuilder} object. In the fix version, the developer caught the \code{RuntimeException} by adding a \code{catch} block with handling code. He also commented on the details of the bug in the handling code. 

In our previous work~\cite{currentwork}, we performed a detailed empirical study on 246 exception bugs and fixes from 10 Android apps. Table \ref{dataset_empirical} shows the number of exception bug fixes for each project in our empirical dataset. The dataset is available at rebrand.ly/ExDataset. We discovered several interesting findings on the nature of exception-related bugs in mobile app development and how app developers fix them. The findings from the study have motivated us in developing recommendation models for exception handling.

\begin{table}[h]
	\centering
	\sf\scriptsize
	\caption{The empirical dataset}
	\label{dataset_empirical}
	\begin{tabular}{llrr}
		\toprule
		\textbf{Name}   & \textbf{Website}  & \textbf{Commits}     & \textbf{Exception Bugs}   \\
		\midrule
		AntennaPod     & antennapod.org    & 3,404                 & 28           \\
		ConnectBot     & connectbot.org  & 1,450                 & 8           \\
		Conversations  & conversations.im   & 3,318                 & 23           \\
		FBReaderJ      & fbreader.org   & 27,944                & 30           \\
		K-9            & k9mail.github.io   & 7,254                 & 31           \\
		MozStumbler    & mozilla.org   & 2,667                 & 7           \\
		PressureNet    & cumulonimbus.ca  & 1,017                 & 11           \\
		Signal   	   & whispersystems.org	  & 2,754                 & 28           \\
		Surespot	   & surespot.me  & 1,590				  & 8			\\
		WordPress	& apps.wordpress.org  & 15,546                & 72          \\
		\midrule
		\textbf{Total} &   & \textbf{66,944} & \textbf{246} \\
		\bottomrule
	\end{tabular}
\end{table}

\subsection{How app developers fix exception bugs?}

First, we focused on how the exception bugs affect the running apps. Exception bugs caused apps crashing in over 80\% of the cases (199/246). Over 13\% (33/246) of the exception bugs caused the app running in an unstable state or some features might not function properly. These effects could lead to negative user experiences and adoptions and cost the development team time and effort to fix the bugs and update the apps. The result suggests that handling exception is an important problem to be solved in app development.

Second, we found out that most of the exception bugs are caused by Android API methods with 51\% (127/246). On the type of exceptions, in 246 exception bugs, there are mostly runtime exceptions with 58.13\% (143/246) of. The result indicates that Android API methods often cause exception bugs while the majority of exceptions in exception bugs are runtime exceptions. The studies in \cite{Kechagia_undocumented, Coelho_unveiling} also show that most methods from Android packages in their stack traces had undocumented runtime exceptions. Due to insufficient documentation, app developers might be unaware of an exception that could occur while using an Android API method, which could lead to a serious bug. The finding suggests that an exception recommendation tool should focus on predicting the occurrences of runtime exceptions for Android API methods. 

\begin{table}[]
	\centering
	\sf\scriptsize
	\caption{Frequency of methods that causes exceptions}
	\label{empirical:table4}
	\begin{tabular}{lr}
		\toprule
		\textbf{OutOfMemoryError}                &  \textbf{19}  \\
		\midrule
		Bitmap.createBitmap             & 5  \\
		Bitmap.decodeResource           & 4  \\
		BitmapFactory.decodeStream      & 4  \\
		BitmapFactory.decodeByteArray   & 4  \\
		Byte.new                        & 2  \\
		\midrule
		\textbf{NumberFormatException}           &  \textbf{17}  \\
		\midrule
		Integer.parseInt                & 8 \\
		Long.parseLong                  & 3  \\
		Integer.valueOf                 & 3  \\
		Long.valueOf                    & 2  \\
		BigDecimal.new                  & 1  \\
		\midrule
		\textbf{ActivityNotFoundException}       &  \textbf{15}  \\
		\midrule
		Context.startActivity           & 4  \\
		Activity.startActivity          & 4  \\
		Activity.startActivityForResult & 3  \\
		Fragment.startActivity          & 3  \\
		BaseActivity.startActivity      & 1  \\
		\bottomrule
	\end{tabular}
\end{table}

Third, we found that there are associations between methods and exception types, as some exceptions have very strong association with specific methods. Table \ref{empirical:table4} shows frequency of methods that causes \code{OutOfMemoryError}, \code{NumberFormatException} and \code{ActivityNotFoundException}. From the table we can see that parsing numbers often throws \code{NumberFormatException}, start an activity could introduces \code{ActivityNotFoundException} and using \code{Bitmap} often causes \code{OutOfMemoryError}. The reverse relationship is also true as all three exceptions are only causes by methods in the table. 

%-----------------------
\begin{figure}[h]
	\centering
	\includegraphics[scale = 0.45]{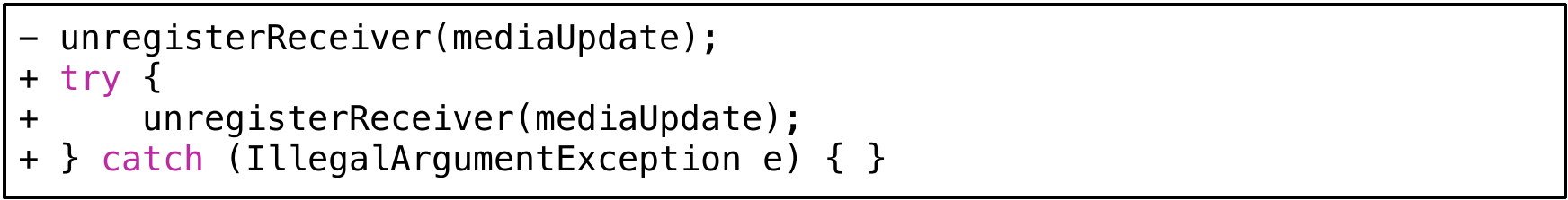}
	\label{example23}
	\caption{An example of swallowing an exception}
	\label{swallowing}
\end{figure}

On handling activities, we found that programmers did not perform any actions, i.e. swallowing exceptions, to handle exception in about 16\% (40/246) bug fixes. An example of swallowing exception is shown in Figure \ref{swallowing}. 
In general, swallowing exceptions is a bad practice, the data and logic in programs might change because of the exception, thus, continue running without modification could introduce new bugs to the program. Our result suggests that there are considerable exception bugs are fixed by swallowing exceptions.
%--------------------------------------------

\begin{figure}[h]
	\centering
	\includegraphics[scale = 0.4]{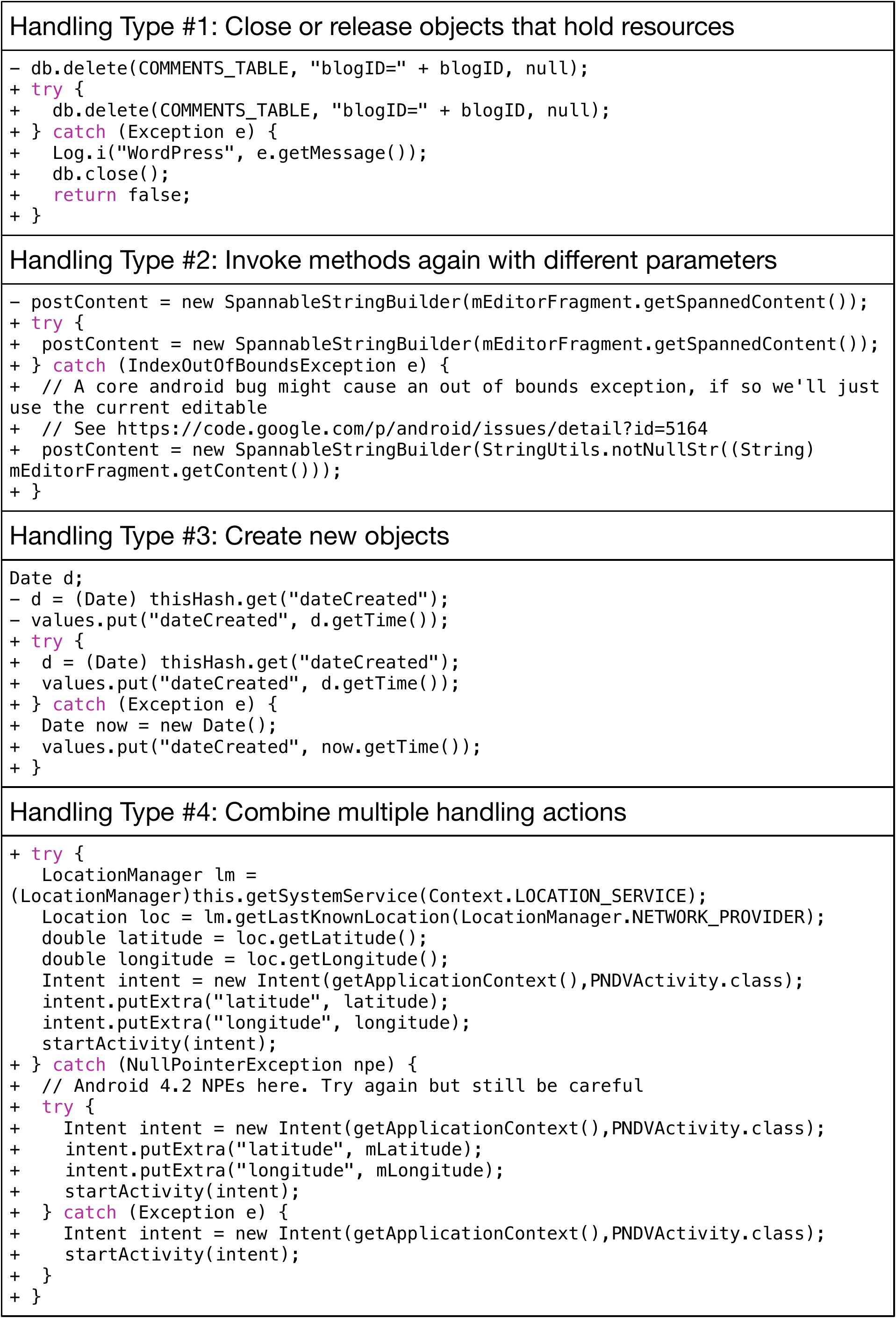}
	\caption{Different types of handling actions}
	\label{example20}
\end{figure}

Programmers invoke method calls or other actions to handle exceptions in 41.86\% (103/246) of bug fixes. We found that about 52\% (54/103) of these bug fixes, programmers used default values to handle exceptions. One example of this handling type is if an exception occurs inside a method that returns an object, in the \code{catch} block of the fix, programmers handle exceptions by returning \code{null}. Programmers invoke method calls in \code{catch} blocks to handle exceptions in the remaining 47 bug fixes. We further found patterns in the remaining 47 bug fixes. The patterns include closing or releasing current used objects that hold resources such as \code{Cursor} or \code{Activity} (10), invoke the method that causes exception again or with different parameters (12), create a new object to replace the object that causes the exception (4), and other actions (21). Examples for each pattern are shown in Figure \ref{example20}. This finding suggests that there are patterns in handling actions of app developers. 

\subsection{How app developers handle exceptions?}

In our previous work~\cite{currentwork}, we also study how professional app developers handle exceptions on high-quality Android apps. We analyzed exception handling code in 4000 top free apps from 36 categories in the Google Play Store. Table \ref{tableiv} shows that statistics of the dataset that we used. Results from the study could be used as guidelines on building exception handling recommendation models.

First, we found that developers spend most of their time to handle runtime exceptions. For example, 7/10 exceptions in top-10 handled exceptions by frequency are runtime exceptions. The result is consistent the study on exception bug fixes as runtime exceptions are most often cause exception bugs. 

\begin{table}[h]
	\centering
	\sf
	\scriptsize
	\caption{Frequency of exception types against a method}
	\label{empirical2:table2}
	\begin{tabular}{lrr}
		\toprule
		\textbf{Activity.startActivityForResult}                & \# & \%\\
		\midrule
		android.content.ActivityNotFoundException             		& 189 &40.6\%\\
		java.lang.Exception           						  		& 164 & 35.2\%\\
		java.lang.ClassNotFoundException      				  		& 33 & 7.1\%\\
		pm.PackageManager.NameNotFoundException   	& 31  &6.7\%\\
		java.lang.SecurityException                               	& 10 &2.1\%  \\
		\midrule
		\textbf{Cursor.moveToFirst}           &  \\
		\midrule
		android.database.sqlite.SQLiteException                		& 4874 & 61.9\%\\
		java.lang.Exception                  						& 2020 & 25.6\% \\
		android.database.SQLException                 				& 264  &  3.3\%\\
		java.lang.Throwable                  						& 180  & 2.2\%\\
		android.database.sqlite.SQLiteFullException                 & 130 & 1.6\%\\
		\bottomrule
	\end{tabular}
\end{table}

Second, we found that there is a high correlation between methods and exception types in terms of co-occurrence. For illustration, Table \ref{empirical2:table2} shows top-5 exception types that co-occur with the methods \code{startActivityForResult} and \code{moveToFirst} by frequency. From the table, we can see that the methods mostly co-occur with the first two exception types in the ranked list. While the general \code{java.lang.Exception} is in the top-2 results, we could infer that each method mostly occurs with the top-1 exception in a rank list, such as \code{ActivityNotFoundException} for the method \code{startActivityForResult}. Based on the observation, one might think we could use the co-occurrence between methods and exception types as a measure to predict which exceptions are likely to occur when using a method. 

Third, we focus on how developers use method calls to handle exceptions. We found that most of the time, developers often only use 1 method call per object in handling code. For example, developers use just 1 method call on the \code{Cursor} object in 95.3\% of all handling code involves the object. In addition, the selection of the method call in handling code for each object is also limited, often falls into 1 or 2 methods. For example, developers often call \code{disconnect} method (50.5\%) or \code{getErrorStream} method (36.9\%) on handling code involves the \code{HttpURLConnection} objects.

\subsection{Summary}

From our study, we found that there is a considerable number of exception-related bugs occurs in app developments. Those bugs often cause serious problems for apps such as crashing or apps running in an unstable state. To fix those bugs, app developers still use bad practices such as swallowing exceptions. In several exception bug fixes, we also found that app developers did not handle exceptions properly. Thus, the results suggest the need for exception support tools to help app developers prevent exceptions from happening and assist developers to handle exceptions correctly. 

We found that most of exception bugs occur by runtime exceptions and developers also spend most of their time to handle runtime exceptions. Thus, we should build models to predict potential runtime exceptions that might occur given a piece of code. Such models are useful to detect potential exception bugs. We also observe there is a high correlation between methods and exception types in terms of co-occurrence. Thus, it suggests that we could build a model to predict potential runtime exceptions using co-occurrences between methods and exceptions collected from a large amount of code.

Our study shows that there are patterns code patterns appear in exception handling code. For example, developers often use four main types of actions to fix exception bugs including close or release objects that hold resources, invoke methods with different parameters, create new objects, and combine multiple actions. Developers also often use specific method calls in handling code. A model that captures these patterns could be useful in recommending exception handling code.

\section{Approach}

This section briefly discusses the points in the design and implementation of our approach on recommending exception handling code patterns. We will start by introducing {\firstmodel}, a fuzzy-based model to rank and suggest exceptions that might occur given a piece of code. Then we describe {\secondmodel}, a machine learning model for recommending exception repairing actions. Finally, we discuss the overall structure of our code recommendation tool for exception handling {\tool} (\textbf{Ex}ception \textbf{Assist}ant) and how {\tool} utilizes the two proposed models to perform its functions.

\subsection{XRank - Ranking Exception Types}

In {\one}, the problem of ranking exceptions is modeled as follows: given a code snippet $C$ that contains a set of method calls $S$, find the exception(s) $E$ with the highest possibility to occur when $C$ is executed. {\one} ranks exceptions $E$  toward a set of method calls $S$ by modeling the correlation/association of $E$ with method calls in $S$. If an exception $E$ has a higher correlation with method calls in $S$, $E$ is considered to have a higher possibility to occur on $C$, and will be rank higher. 

To model correlation/association between method calls and exceptions, {\one} utilizes fuzzy set theory \cite{klir_fuzzy}. It defines a fuzzy set of potential exceptions toward a method as follows.

\begin{definition}[Potential Exception]
	For a specific method call $m$, a fuzzy set $C_m$, with an associated membership function $\mu_m()$, represents the set of potential exceptions toward $m$, i.e. exceptions that are highly correlated with $m$.
\end{definition}

Fuzzy set $C_m$ is determined via a membership function $\mu_m$ with values in the range $[0,1]$. For an exception $E$, the membership score $\mu_{m}(E)$ determines the certainty degree of the membership of $E$ in $C_m$, i.e. how likely $E$ belongs to the fuzzy set $C_m$. In this context of {\one}, $\mu_m(E)$ represents the degree of association between $E$ and $m$. $\mu_m(E)$ is also determines the ranking of $E$ toward $m$. If $\mu_m(E) > \mu_m(E')$ then $E$ is considered higher correlated to $m$ to $E'$. The membership score is computed as follows.

\begin{definition} [Membership Score]
	The membership score $\mu_{m}(E)$ is computed as the correlation between the set $D_{m}$ represents usages of method $m$, and the set $D_{E}$ represents usages of exception $E$:
	\begin{equation} \label{eq:1}
	\mu_{m}(E) = \frac{|D_m \cap D_{E}|}{|D_m \cup D_E|} = \frac{n_{m, E}}{n_m + n_E - n_{m, E}} 
	\end{equation}
\end{definition}
where, $n_m$ is the number of usages of method $m$, $n_E$ is the number of usages of exception $E$, and $n_{m, E}$ is the number of times that exception $E$ occurs on code snippets that contains method $m$. As the formula \ref{eq:1}, the value of  $\mu_{m}(E)$ is between $[0,1]$. If $\mu_{m}(E) = 1$, then $E$ always occurs on the code snippets that contain method $m$ in the codebase, thus, given a code snippet contains $m$, it is very likely to catch exception $E$. If $\mu_{m}(E) = 0$, it means that the exception never occurs on code snippets that contains $m$, thus, given a code snippet contains $m$, it is unlikely to catch exception $E$. In general, the more frequently exception $E$ occurs on code snippets contains $m$, the higher value of $\mu_{m}(E)$. 

%Using membership score $\mu_{m}(E)$ as Formula \ref{eq:1}, {\one} will also rank higher exception $E$ that often co-occurs only with some method calls (e.g. \code{SQLiteException}) over general exception types such as \code{java.lang.Exception}. That is, if both $E$ and $E'$ have similar frequency to occur on code snippets that contains method call $m$, or $n_{m, E} \approx m_{m, E'}$, but $E'$ is also used to catch with several methods other than $m$ while most of the time $E$ is only used to catch on $m$. Thus, the value of $n_{E'}$ will be much larger than the value of $n_{E}$. Therefore, {\one} will give $E$ higher score compared to $E'$. In general, {\one} will rank higher exception $E$ that is often used to catch code snippets that contains $m$ (large $n_{m, E}$) or is designed to catch on only some method calls including $m$ (large $n_{m,E}/n_{E}$). 

Based on the definition of potential exceptions toward a method as a fuzzy set, {\one}  defines potential exceptions toward a set of methods using the union operation of fuzzy set theory as follows.

\begin{definition}
	Given a set of method calls $S$, a fuzzy set $C_S$, with an associated membership function $\mu_S()$, represents the set of potential exceptions toward $S$, i.e. the exceptions that are highly correlated with methods of $S$. $C_S$ is computed as the union of the fuzzy sets for method calls in $S$
	\begin{equation}
	C_S = \displaystyle\bigcup_{m \in S} C_m 
	\end{equation} 
\end{definition}

The membership score of an exception $E$ with respect to the union set $C_S$ is computed as follows.

\begin{definition}
	The membership score $\mu_{S}(E)$ is calculated as the combination of the membership scores $\mu_{m}(E)$ of its associated method $m$:
	\begin{equation} \label{eq:3}
	\mu_{S}(E) = 1 - \prod_{m \in S} ( 1-\mu_{m}(E)) 
	\end{equation}
\end{definition}

$\mu_{S}(E)$ represents the correlation of exception $E$ toward a set of method calls $S$. As the equation, we see that the value of $\mu_{S}(E)$ is also between $[0,1]$ and represents the likelihood in which the exception $E$ belongs to the fuzzy set $C_S$, i.e. the set of potential exceptions for the set of method calls $S$. $\mu_{S}(E) = 0$ when all $\mu_{m}(E) = 0$, which means that exception $E$ never occurs on any code contains a method in $S$. Thus, {\one} considers that exception $E$ is unlikely to occur on the code contains $S$. If there is any method $m$ is $S$ with $\mu_{m}(E) = 1$, then $\mu_{S}(E) = 1$, or {\one} considers that exception $E$ is very likely to occur on the code contains $S$ as $E$ always occurs on code contains a method $m$ in $S$. In general, the more methods in $S$ with high $\mu_{m}(E)$ values, the higher $\mu_{S}(E)$ is, or $E$ is more likely to occur on the code contains $S$. 

%Computing correlation score $\mu_{S}(E)$ as Formula \ref{eq:3} has several advantages as it allows {\one} to take into consideration of co-occurring method calls. If both of methods $m$ and $m'$ have high correlations with exception $E$, the set of method $S$ contains $m$ and $m'$ is expected to have higher correlation with $E$ than each of its method call since $m$ and $m'$ co-occur will increase the likelihood of catching exception $E$. This observation is true using this formula. For example, assume that $\mu_{m}(E) = 0.75$ and $\mu_{m'}(E) = 0.65$. Then, $\mu_{S}(E) = 1 - (1-0.75)\times(1-0.65) = 0.9125$, which is higher than each of $\mu_{m}(E)$ and $\mu_{m'}(E)$. $\mu_{S}(E)$ is also not affected much if it contains a method $m'$ that has low correlation with exception $E$. For example, let assume that $\mu_{m}(E) = 0.75$ and $\mu_{m'}(E) = 0.1$. Then, $\mu_{S}(E) = 1 - (1-0.75)\times(1-0.1) = 0.775$, which is not much higher than each of $\mu_{m}(E) = 0.75$.

After computing catching correlation scores $\mu_{m}(E)$ of exception $E$ and each method call $m$ in $S$. {\one} uses Formula \ref{eq:3} to compute correlation score $\mu_{S}(E)$ of $E$ and $S$, then, it ranks and recommends exceptions with top scores as the most likely exception to catch for $S$.

\subsection{XHand - Exception Handling Models}

We define the problem of recommending repairing method call sequence as follows: given a set of method calls $S_{n}$ of an object of type $T$ in a \code{try} block, recommend the most reasonable repairing method call sequence $S_{h}$ in the corresponding \code{catch} block. 

{\two}'s approach for this problem is based on the idea of statistical models. If we could learn patterns from the history of handling method calls in $S_{n}$, we could use these patterns to suggest $S_{h}$. 

\begin{figure}[h]
	\centering
	\includegraphics[scale = 0.4]{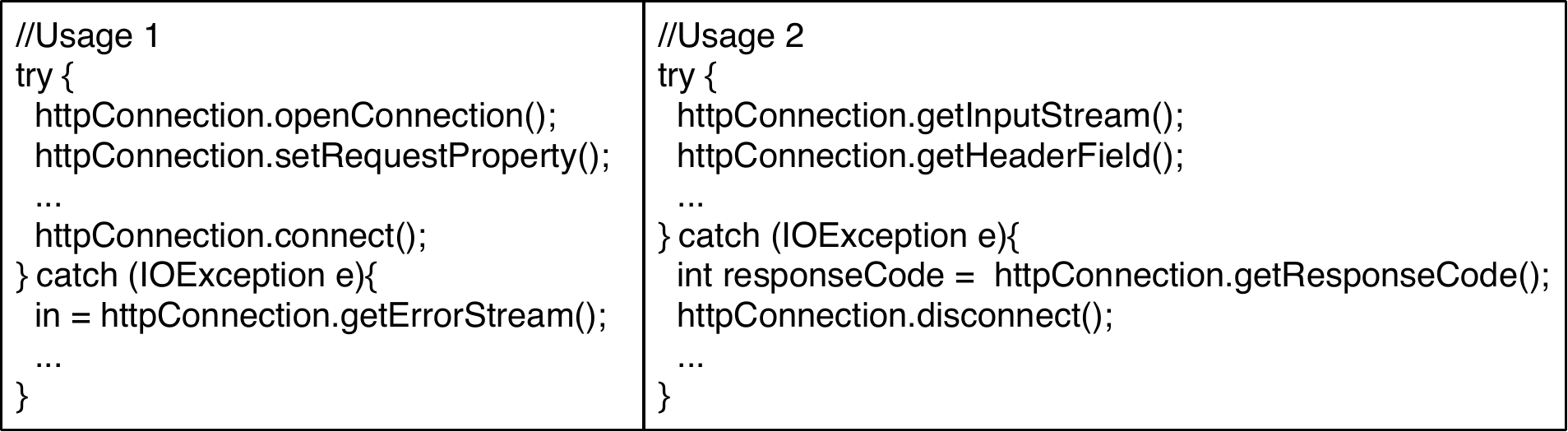}
	\caption{Handling exceptions of \code{HttpURLConnection} object}
	\label{code:http}
\end{figure}

One observation when developing {\two} is that handling method sequence $S_{h}$ depends on the current context of the object. While the current context of the object could be represent as by the set of method calls $S_{n}$ in try block. For example, consider two usages of a \code{HttpURLConnection} in Figure \ref{code:http}. 

In the first usage, the exception may occur while using connecting to a server via \code{connect} method. In this situation, we would want to get the error stream by calling \code{getErrorStream} to get information about the error for further processing. While in the second usage, the object is already connected to the server but the read timeout expires causing an exception. In this situation, the exception handler, we would want to get the response code and disconnect from the server.

This observation suggests that predicting the first repairing method call could be formalized as a multi-class classification problem \cite{statisticallearning}. In this problem, an instance is a set of method calls appear in \code{try} block. It represents the context of an object in the \code{try} block in which an exception could occur. The predicting label is the first repairing method call in \code{catch} block. It represents the starting action of repairing. Thus, to predict the first repairing method, firstly, {\two} learns a multi-class classification model and uses that model for recommendations.

{\two}'s recommendation for remaining handling method calls  is developed based on $n$-gram model \cite{hindle_naturalness, ngram}. $n$-gram model is one type of statistical language models that learns all possible conditional probabilities
\begin{equation}
P(m_{i}|m_{i-n+1}...m_{i-1}) \nonumber
\end{equation}
where $m_i$ is the current word and $m_{i-n+1}...m_{i-1}$ is the sub-sequence of $n-1$ prior words. This is the probability that $m_i$ occurs as the next word of $m_{i-n+1}...m_{i-1}$. Using the chaining rule, we can use an $n$-gram model to compute the generating probability of any given sentence $s = (m_1...m_n)$:
\begin{equation}
P(s) = \prod_{i=1}^{n} P(m_{i}|m_{i-n+1}...m_{i-1}) 
\end{equation}
In $n$-gram model, the probability of the next word $m_{i}$ only depends on the previous $n-1$ words. When training $n$-gram model, the conditional probabilities  $P(m_{i}|m_{i-n+1}...m_{i-1})$ are estimated by counting the number of occurrences of $n$-gram and $(n-1)$-gram in the training data:
\begin{equation}
P(m_{i}|m_{i-n+1}...m_{i-1}) = \frac{N(m_{i-n+1}...m_{i})}{N(m_{i-n+1}...m_{i-1})} 
\end{equation}

In the context of learning exception repairing patterns, a repair method call sequence is considered as a sentence and the first handling method call is considered as the first word of a sentence. 

After training $n$-gram, given $i-1$ previous method calls in exception handler, the recommendation for method calls at position $i$ of exception handler is computed follows. For each method $m_i$, the $score$ of $m_i$ is approximated by the conditional probability of $m_i$ given previous $n-1$ words:
\begin{equation}
score(m_i) = P(m_{i}|m_{i-n+1}...m_{i-1}) 
\end{equation}
Sorting $m_i$ by $score$, we will have a ranked list of method call recommendation where method call at higher position in the list has higher probability of calling.

\subsection{Design Overview}
\begin{figure}[h]
	\centering
	\includegraphics[scale = 0.35]{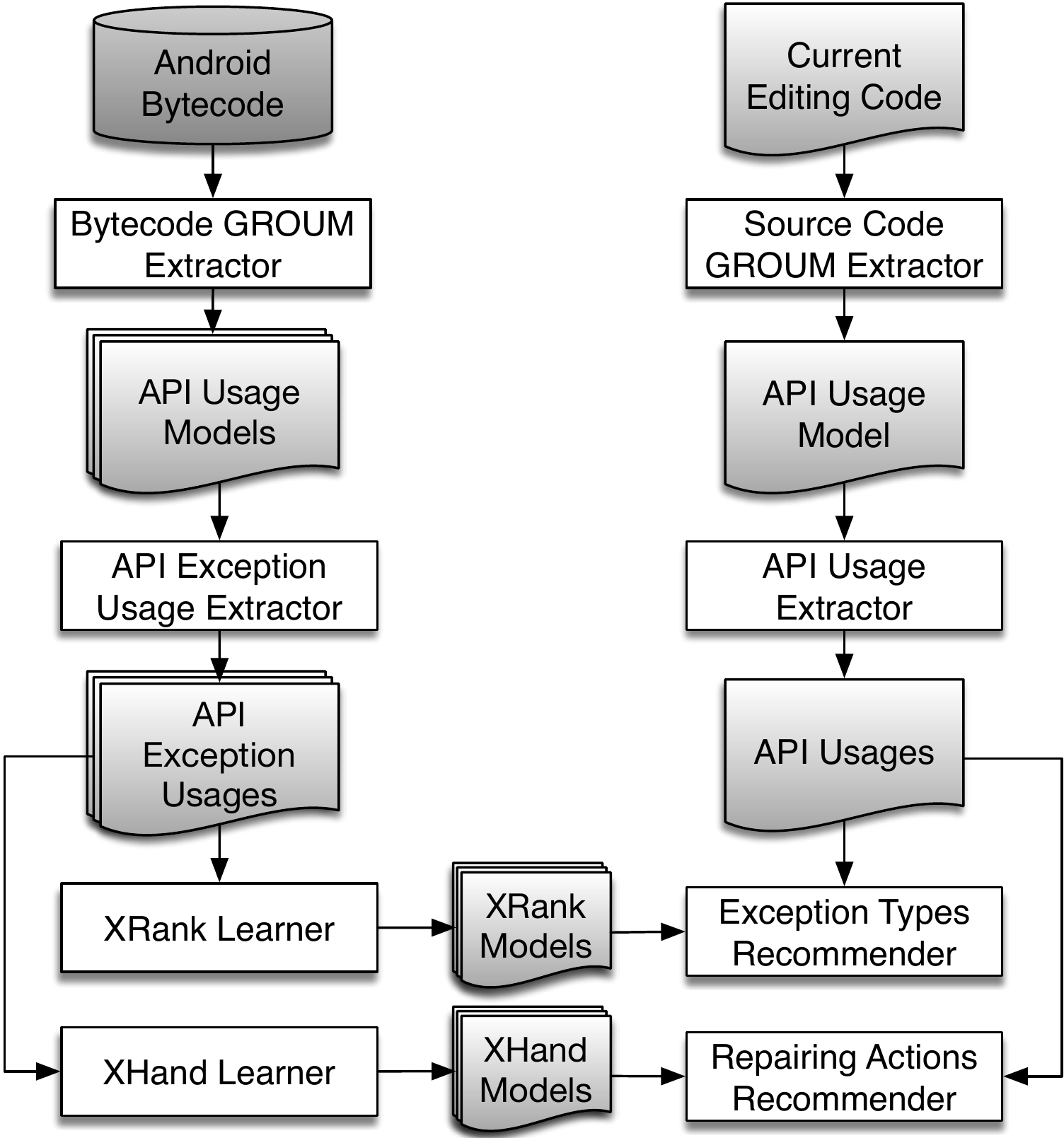}
	\caption{Design Overview of {\tool}}
	\label{fig:tool}
\end{figure}

Figure \ref{fig:tool} shows the design overview of {\tool}. It has five major components including two modules to extract API usage models (represented as graph-based object usage models - GROUM~\cite{groum}) from source code and bytecode, a module to extract API exception usages from those usage models, two modules to train {\firstmodel} and {\secondmodel} from those extracted data, and two modules that use the trained models to recommend exception types and exception repairing actions. Let us describe those modules in more details.

\noindent\textbf{1. GROUM Extractor.} {\tool} uses {\groum} (Graph-based Object Usage Model)~\cite{groum} to represent the raw API usages in source code and bytecode. To collect training data, {\tool} has a module to extract GROUMs from bytecode of existing Android apps. It also has a similar module to extract GROUMs from the code being written, which are used for its two tasks of recommending exception types and repairing actions. The extracting algorithms could be found at~\cite{nguyen_salad,nguyen_droidassist,groum}\\
\textbf{2. API Exception Usage Extractor.} Because {\firstmodel} and {\secondmodel} are learned from usages of API objects and method calls in exception handling code, {\tool} has a module named \code{API Exception Usage Extractor} to extract from those usages from GROUMs. For each sub-graph of a {\groum} represents an exception handling case (i.e. a \code{try-catch} block), this module traverses through the sub-graph to extract API method calls in the \code{try} block, the catching exception type, handling API method calls in the \code{catch} block. The temporal order and data dependency between those API method calls are also extracted. The module then stores that information as an API exception usage.\\
\textbf{3. {\firstmodel} and {\secondmodel} Learners.} These modules are responsible for training {\firstmodel} and {\secondmodel} models from the extracted API exception usages. The {\firstmodel} learner computes membership scores for each pair an API method call $m$ and an exception type $E$. It invokes counting usages of $m$, $E$, and the number of times both $m$ and $E$ appear in an exception handling code. Each {\secondmodel} model is a multi-class prediction model for an object type. Given a set of method calls $S_n$ in a \code{try} block and the catching exception $E$, the model predicts the repairing method call sequence $S_h$. {\secondmodel} is learned from a dataset contains a set of pairs $(S_n, S_h)$. The training dataset could be constructed from the API exception usages extracted by the previous module.\\
\textbf{4. Exception Types Recommender.} This module provides the recommendations on unchecked exception types for a selected piece of code. Its input is a set of API method calls appeared in the selected code. The module then utilizes {\firstmodel} to compute confident scores for each exception types against the set of API method calls. If the confident score of an exception type is greater than a threshold, the exception type is considered likely to occur and it will be included in the rank list of recommendation. The threshold is computed based on the distribution of confident scores of all set of method calls and exception types used for training of {\firstmodel}. Based on our experiment, we currently use the 25\% percentile as the threshold. That means if the {\firstmodel} is trained with 100 pair of API method calls and exception types, and 25 of them have probabilities less than 0.1, this value is chosen as the threshold.\\
\textbf{5. Repairing Actions Recommender.} This module provides the recommendations for repairing actions in exception handling code. The input is a set of API method calls in a \code{try} block $S_n$ and the catching exception $E$. The module then groups API method calls by objects. All API method calls in a group of an object have data dependency with that object. Static methods are grouped by object types. The module uses the corresponding {\secondmodel} model to predict repairing actions for each object given a set of API method calls. Finally, it combines all the predict repairing actions to produce the final recommendation.

\section{Tool Introduction}
In this session, we introduce the usages of our code recommendation tool for exception handling, {\toolex}. The tool is available at: https://rebrand.ly/exassist. {\toolex} predicts what types of exception could occur in a given piece of code and recommends proper exception handling code for such an exception. When requested, it will add such code into the given piece of code. {\toolex} is released as a plugin of IntelliJ IDEA and Android Studio, two popular IDEs for Java programs and Android mobile apps. After installation, it is incorporated with the IDE and users can invoke it directly via shortcut key \code{Ctrl + Alt + R} or via the menu bar. Let us present {\toolex}'s main functionality via two usage scenarios. 

% \begin{figure*}[t]
% 	\centering
% 	\includegraphics[scale = 0.4]{figures/first_usage}
% 	\caption{Recommending Exception Types  by {\toolex}}
% 	\label{first:first}
% \end{figure*}

\begin{figure}[h]
	\centering
	\includegraphics[scale = 0.5]{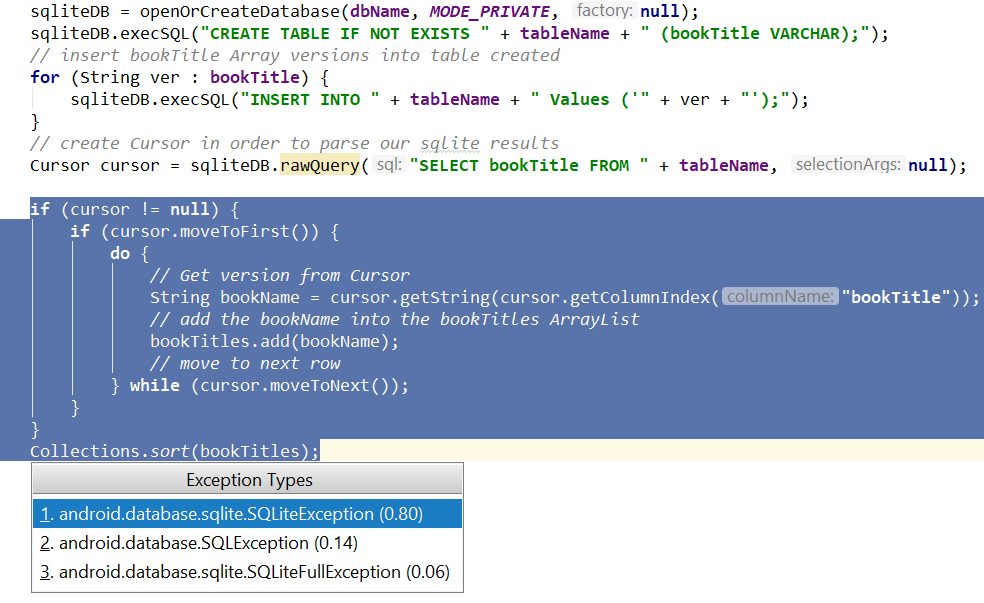}
	\caption{Recommending Exception Types  by {\toolex}}
	\label{first:first}
\end{figure}

\noindent\textbf{Recommending Exception Types.} Assume a developer is writing code to open and get data from a database (Figure \ref{first:first}). She first opened a \code{SQLiteDatabase} and linked it to variable \code{sqliteDB}. She then performed some \code{SQL} queries on \code{sqliteDB} to update the database. Next, she created a database query that returns a \code{Cursor} object and navigated through that \code{Cursor} object to retrieve the data to a list named \code{bookTitles}. The developer is aware that the code is dealing with database and \code{Cursor} objects might throw unchecked exceptions at runtime, but she might be unsure whether to catch exceptions on the code she wrote and which type of exception to be caught. The built-in exception checker in Android Studio only supports adding checked exceptions, thus, does not help her to make appropriate action in this case.

{\toolex} aims to support the developer to make decisions whether or not to add a \code{try-catch} block and what type of exception to catch. The developer invokes {\toolex} by first selecting the portion of code that she wants to check for an exception then pressing \code{Ctrl + Alt + R}. Figure \ref{first:first} shows a screenshot of Android Studio with {\toolex} invoked for the portion of code that using the \code{Cursor} object for reading data from the database. As seen, {\toolex} suggests that the code is likely to throw an unchecked exception. It also displays a ranked list of unchecked exceptions that could be thrown from the current selecting code. Each unchecked exception in the ranked list has a confident score represents how likely the exception will be thrown from the code. The value for confident scores is between 0 and 1. The higher the value of the confidence score, the higher the likelihood the exception type is thrown. In this example, \code{SQLiteException} has the highest score of 0.80. If the developer chooses that exception type, the currently selected code will be wrapped in a \code{try-catch} block with \code{SQLiteException} in the \code{catch} expression.

%  \begin{figure}[h]
%  	\centering
%  	\includegraphics[scale = 0.5]{figures/second_usage}
%  	\caption{Recommendation for the usage of object \code{sqliteDB}}
%  	\label{first:second}
%  \end{figure}

% {\toolex} uses the context of current selecting code to infer whether or not adding exception handling code and the type of the exception. For example, in Figure \ref{first:second}, the context changes as the developer selects the portion of code for opening and querying on the \code{SQLiteDatabase} object. Thus, {\toolex} updates the recommendation list with \code{SQLException} has the highest confident of 0.81, which is highest among all other exception types.

% {\toolex} could provide recommendations for a selected portion of the code includes one or multiple method calls. Additionally, {\toolex} could also recommend not to add \code{try-catch} block if it infers that the selected code is very unlikely to throw an unchecked exception. For example, if the developer selects the statement \code{bookTitles.add(bookname);} and queries {\toolex}, the tool will return an empty list of exceptions as it is very unlikely the selected method throws exceptions when it is executed.

\begin{figure}[h]
	\centering
	\includegraphics[scale = 0.5]{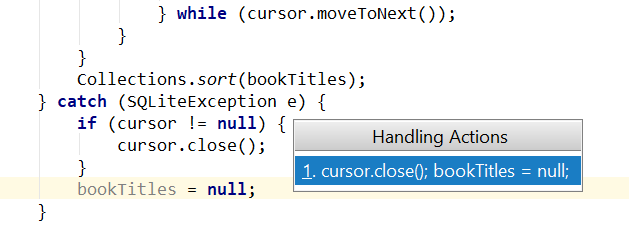}
	\caption{Recommending repairing actions}
	\label{second:first}
\end{figure}

%  Handling exception situations and executing necessary recovery actions are important as it could help apps continue to run properly when an exception occurs. For example, when an app reuses resources such as database connections or files, the app should release the resources if an exception is thrown. Failing to release the resources can cause performance degradation and lead to unexpected behaviors of the app. {\toolex} is also designed to recommend such repairing actions in the exception handling code based on the context in the \code{try} block. 

\noindent\textbf{Recommending Exception Repairs.} Figure \ref{second:first} demonstrates an usage of {\toolex} in recommending exception repairing actions. After adding a \code{try-catch} block with \code{SQLiteException} for the code in the previous scenario, the developer wants to perform recovery actions. To invoke {\toolex} for this task, she moves the cursor to the first line of the \code{catch} and presses \code{Ctrl + Alt + R}. {\toolex} then will analyze the context of the code and provide repairing actions in the recommendation windows. In the example, {\toolex} detects that the \code{Cursor} object should be closed to release all of its resources and making it invalid for further usages. It also suggests to set \code{bookTitles} equals \code{null} to indicate the error while collecting data from \code{cursor}. If the developer chooses the recommended actions, {\toolex} will generate the code in the \code{catch} block as in the Figure \ref{second:first}. To save space, we show both the recommendation and generated code in the same window.

\section{Evaluation}
We performed several experiments to evaluate the effectiveness of our techniques on learning and recommending exception handling code patterns on Android and Java APIs. All experiments are executed on a computer 64-bit Windows 10 with Intel Core i7 3.6Ghz CPU, 8GB RAM, and 1TB HDD storage. 
\subsection{Data Collection}
The dataset used in our evaluation is summarized in Table \ref{table:data_collection}. In total, we downloaded and analyzed 4000 top free apps from 36 categories in the Google Play Store. To ensure the quality of training data, when crawling apps, the app extractor only downloaded apps has an overall rating of at least 3 (out of 5). This filtering is based on the assumption that the high-rating apps would have a high quality of code, and thus, would have better exception handling mechanism. 

%ok
Since Android mobile apps are distributed as \code{.apk} files, our app extractor unpacked each \code{.apk} file and kept only its \code{.dex} file. The total storage space for the \code{.dex} files of all downloaded apps are around 19.9 GB. After parsing those \code{.dex} files we obtained about \textbf{13 million} classes. We next developed a bytecode analyzer that analyzed each class and looked for all methods in the class to build {\groum} models. Since an Android mobile app is self-contained, its \code{.dex} files contain bytecode of all external libraries it uses. That leads to the duplication of the bytecode of shared libraries. Thus, our bytecode analyzer maintains a dictionary of the analyzed methods, thus, is able to analyze each method only once. In the end, we analyzed over \textbf{16 million} methods which have in total nearly \textbf{340 million} bytecode instructions.

\begin{table}[h]
	\centering
	\caption{Data Statistics}
	\label{table:data_collection}
	\scriptsize\sf
	\begin{tabular}{lr}
		\toprule
		\textbf{Data Collection} & \\
		\midrule
		Number of apps                     & 4000 \\
		Number of classes                         & 13,969,235 \\
		Number of methods                         & 16,489,415 \\
		Number of bytecode instructions & 341,912,624 \\
		Space for storing .dex files				  & 19.9 GB \\
		\midrule
		\textbf{Recommending Exception Types} & \\
		\midrule
		Number of exception types                   & 261 \\
		Number of API methods                     & 64,685 \\
		Number of uncaught method sets                         & 16,489,415 \\
		Number of caught method sets                       & 549,786 \\
		\midrule
		\textbf{Recommending Repairing Methods} & \\
		\midrule
		Number of API object types                     & 360 \\
		Total number of pairs $(S_{n}, S_{h})$                         & 187,994 \\
		Average number of pairs per object type                       & 522.20 \\
		\bottomrule
	\end{tabular}
	\label{tableiv}
\end{table}

\subsection{Recommending Exception Types}
\begin{figure*}[t]
	\centering
	\begin{minipage}{0.3\textwidth}
		\centering
		\includegraphics[scale = 0.08]{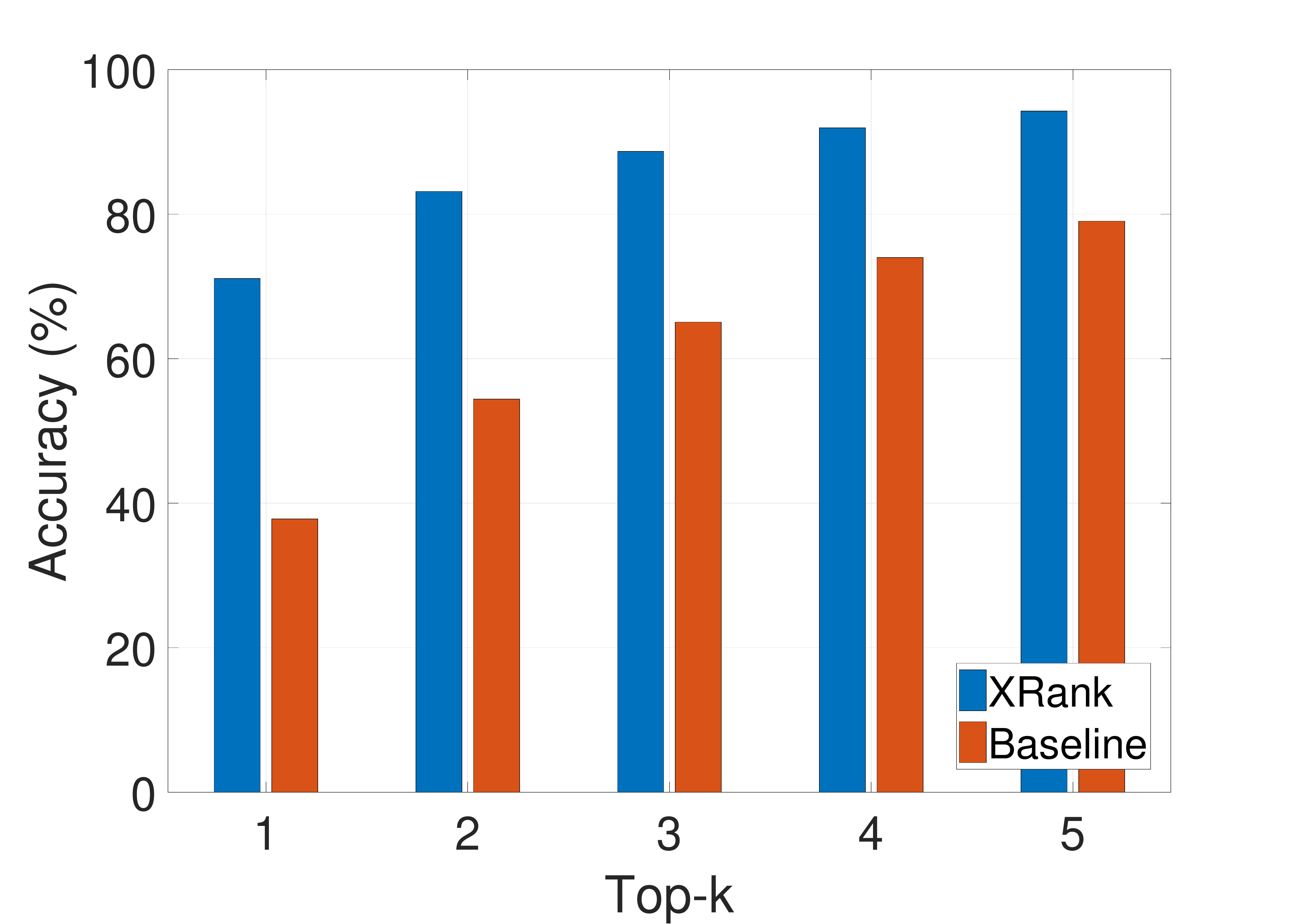}
		\caption{The accuracy of {\one}}
		\label{xrank}
	\end{minipage}
	\hfill
	\begin{minipage}{0.3\textwidth}
		\centering
		\includegraphics[scale = 0.08]{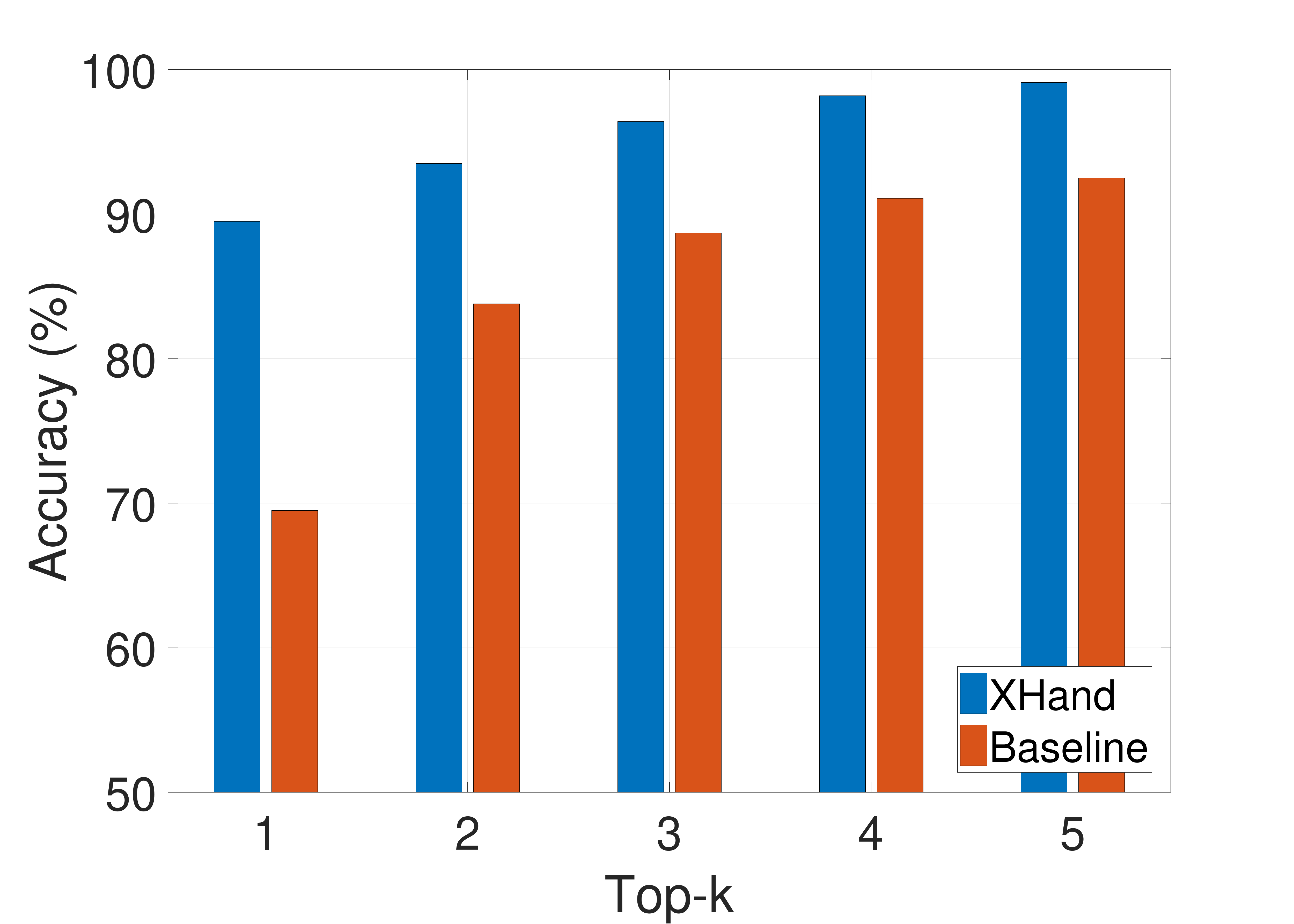}
		\caption{The accuracy of {\two}}
		\label{xhand}
	\end{minipage}
	\hfill
	\begin{minipage}{0.3\textwidth}
		\centering
		\includegraphics[scale = 0.08]{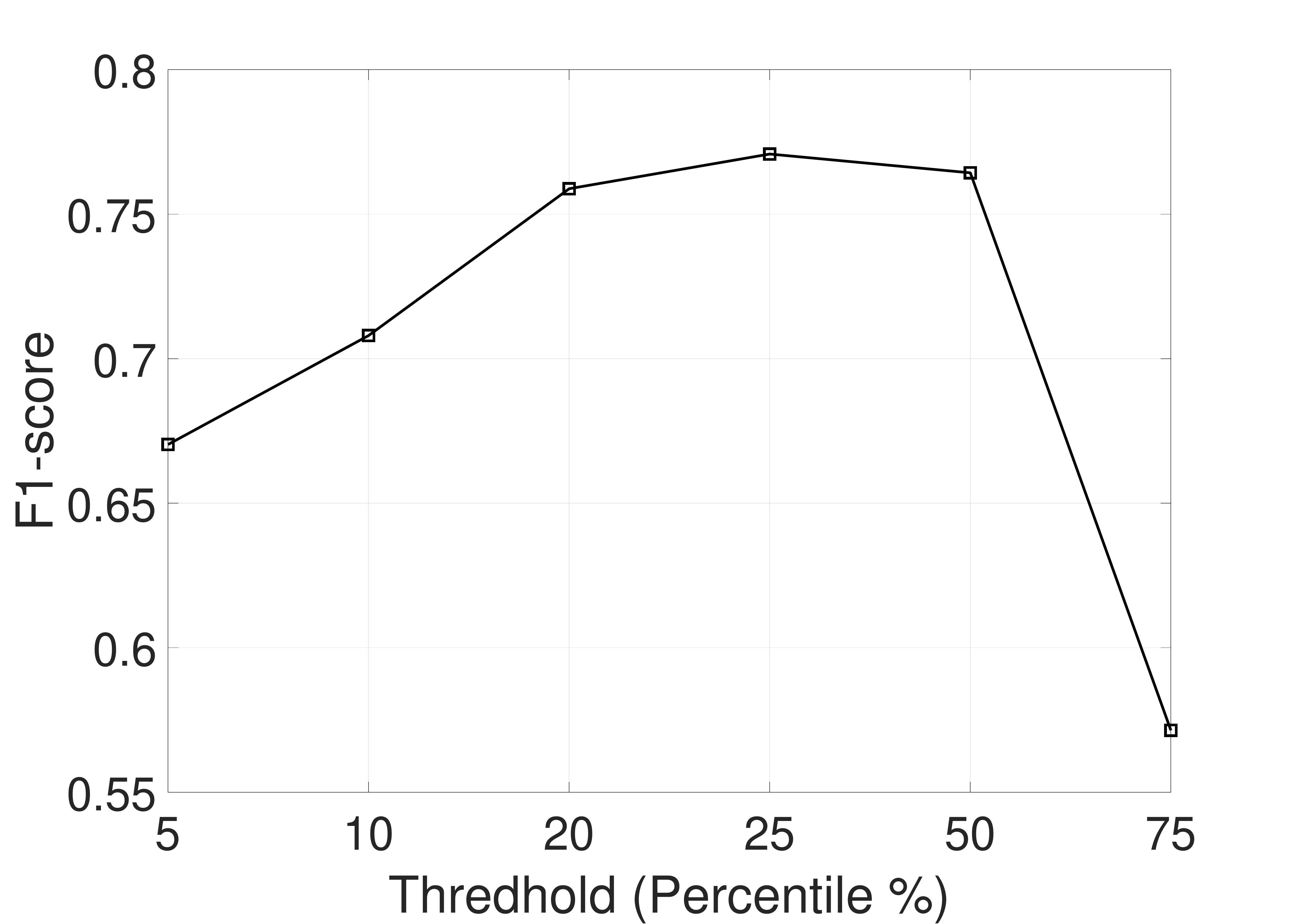}
		\caption{The F1-score by threshold}
		\label{xrank2}
	\end{minipage}
\end{figure*}

%ok
For training and testing {\one}, the bytecode analyzer extract all caught and uncaught method sets from {\groum}. A caught method set is a set of method calls that appear in the same \code{try} block, and is associated with an exception in the \code{catch} block. An uncaught method set is the set of all methods calls appeared in the same method implementation that is not in any \code{try } blocks. Uncaught method sets represent usages of methods that are not associated with exceptions and vice versa.

Table \ref{table:data_collection} shows the statistics of the dataset that we extracted to evaluate {\one}. Overall, the dataset is sufficient for the evaluation. We collected over 16 millions uncaught method sets and roughly 550 thousands caught method sets. There are over 64 thousand Android API methods and a total of 260 exception types. 

We evaluated the performance of {\one} on the task of recommending exceptions. Given a set of methods $S$ in the testing set, let also assume that $S$ is caught with an exception $E$ in the code, {\one} will return a rank list of exception types $R$, each exception type $r_i$ in $R$ is associated with a score represents the possibility of catching $r_i$ on $S$. To evaluate {\one}, we find the position of $E$ in $R$, if $E = r_k$ we consider it as a hit at top-$k$. We then evaluate for all method sets that have been caught exceptions in testing sets. Finally, we calculate the top-$k$ accuracy as the ratio of the total hit at $k$ over the size of the testing set.

We also used 10-fold cross-validation to evaluate {\one}. The dataset of apps was divided into 10 folds. At each iteration, we trained {\one} with nine folds then tested {\one} on the remaining fold. Finally, we averaged the results of all iterations to produce the final top-$k$. 

To measure the effectiveness of {\one}, we compared it with a baseline based on the frequency of catching exception types. Given a caught method set $S$, we built a list of exception types $R$ such as each exception type $r_i \in R$ . The list $R$ is ranked based on the frequency of each exception types appeared in the training set. The assumption is if an exception type appeared more frequent would have a higher rank.

Figure \ref{xrank} shows the experiment results on recommending exceptions to catch. From the figure, we can see that both {\one} and the baseline have consistent accuracy. More importantly, {\one} recommends exception with very high levels of accuracy. For example, it has top-1 accuracy of 70\%, top-3 accuracy of 87\%. Top-5 accuracy of {\one} model approaches 94\%. In addition, {\one} significantly outperforms the baseline. For example, the corresponding top-1 and top-3 accuracy of n-gram model are 37\% (52\% lower) and 55\% (63\% lower). Overall, {\one} has a significant improvement over the baseline. The result shows that the context (method calls in \code{try} blocks) which is captured by {\one} plays an important role in the accuracy of the recommendation model.

\subsection{Recommending Repairing Method Calls}

For evaluating {\two}, the bytecode analyzer extracted all pairs $(S_{n}, S_{h})$ where $S_{n}$ is the set of method calls in \code{try} blocks and $S_{h}$ is the  repairing method call sequence in the corresponding \code{catch} block. Because {\two} is learned for each object type, thus, all method calls in both $(S_{n}, S_{h})$ belong to a same class. 

We performed an experiment to evaluate the accuracy of {\two} model in recommending repairing method call sequences. We chose the task of recommending the next repairing method call. Given the set of method calls $S_{n}$ in a \code{try} block and $i-1$ previous repairing method calls in the exception handler, the {\two} model is expected to recommend the most probable next method call at position $i$ of the exception handler. This recommendation task has been widely used in software engineering research to evaluate code completion models \cite{ngram, gralan, nguyen_salad}.

In this type of experiment, the {\two} model predicts method calls at all positions of every exception handling method sequence $S_{h}$ in the testing set. {\two} utilizes a multi-class classification model to predict the first method in the handling sequence. For a method call $c_i$ at position $i$ of $S_{h}$, we use the {\two} model to infer the top-$k$ most probable method calls $T_k = \{h_1, ..., h_k\}$. If $c_i$ is in $T_k$, we consider it as hit, i.e. an accurate top-$k$ recommendation. The top-$k$ accuracy is the ratio of the total hits over the total number of recommendations. 

All {\two} models are trained using J48 decision tree \cite{statisticallearning} and 2-gram model with Witten-Bell smoothing technique \cite{ngram, nguyen_salad}. Similar to {\one} , our experiment for {\two} is a 10-fold cross validation. We only test API object types that have more than or equals to 10 pairs of $(S_{n}, S_{h})$. As shown in the Table \ref{table:data_collection}, we performed the experiment on 360 object types. On average each object type has 522 pairs of $(S_{n}, S_{h})$.

We compared {\two} with a baseline model based on just the frequency of repairing method calls. At each position $i$, the baseline model recommends the rank list of repairing method calls based on their frequency. The method calls at previous positions are removed from the rank list when providing a recommendation.

Figure \ref{xhand} shows the top-5 accuracy of {\two} and the baseline on recommending repairing method call sequences. From the figure, we can see that {\two} recommends method calls with very high levels of accuracy. For example, it has top-1 accuracy of 89\%, top-2 accuracy of 92\%, top-3 accuracy of 96\%. Top-5 accuracy of {\two} approaches 100\%. Additionally, {\two} outperforms the baseline significantly. The baseline achieves 70\% (23\% lower) in top-1 accuracy, 83\% (11\% lower) in top-2 accuracy, and 89\% (8\% lower) in top-3 accuracy.

\subsection{Empirical Evaluation}
%We have developed {\tool} as a code recommendation tool for exception handling. {\tool} is released as a plugin of IntelliJ IDEA and Android Studio, two popular IDEs for Java programs and Android mobile apps. {\tool} predicts what types of exception could occur in a given piece of code and recommends proper exception handling code for such an exception. When requested, it will add such code into the given piece of code.
In the section, we present our evaluation on the effectiveness of {\tool} on detecting and fixing real exception bugs. 

\subsubsection{Detecting Exception Related Bugs}
With the usage described above, we can see that {\tool} could be applied to detect real exception related bugs. In particular, we focus on exception bugs. To evaluate {\tool}, we manually collected 128 exception bug fixes from the 10 open-source Android projects in our empirical dataset. Each exception bug fix is a fix for an exception bug in which developers add a \code{try-catch} block to handle exceptions.  To measure the recall in detecting exception bugs, we also collected of 128 code snippets that are very unlikely to throw any exceptions. Each code snippet appears in a project of our empirical dataset and have not be wrapped or contains any \code{try-catch} blocks. It also has not been changed and modified through the development process of the project. The code snippets are considered as \textit{negative} examples. Overall, the dataset contains 256 samples. For each sample, we use {\tool} to recommend whether or not adding a \code{try-catch} block on the code. {\tool} should recommend adding a \code{try-catch} block for an exception bug and recommend not adding any \code{try-catch} blocks on a \textit{negative} example. We then measure the accuracy of {\tool} in this task using F1-score \cite{statisticallearning}.

\begin{table}[h]
	\centering
	\sf
	\caption{Result in detecting exception bugs}
	\label{result:xrank}
	\begin{tabular}{lr}
		\toprule
		Exception bugs                    & 128            \\
		Recommended by {\tool}                & 116              \\
		Top-1 accuracy   & 86              \\
		Top-2 accuracy         & 96              \\
		Top-3 accuracy  & 104            \\
		\bottomrule
	\end{tabular}
\end{table}

% \begin{figure}[h]
% 	\centering
% 	\includegraphics[scale = 0.45]{empirical_examples/exassist_1}
%     \caption{An example of potential incorrect exception handling}
% 	\label{empirical:example6}
% \end{figure}

Figure \ref{xrank2} shows the F1-score of {\tool} by the selection threshold used to determine whether to recommend adding a \code{try-catch} block to a sample. If the threshold is small ($5\%$ or $10\%$), {\tool} has a tendency to recommend adding exception handlers (high precision but low recall), thus, the F1-score is low. Conversely, if the threshold is big ($75\%$), {\tool} is likely to recommend not adding \code{try-catch} blocks, which also yields a low F1-score. From the figure, we can see that the F1-score yields high values in the range from $20\%$ to $50\%$ and peaks at $25\%$. Thus, we selected $25\%$ as the selection threshold to use in further evaluation of our tool.

Next, for all exception bugs detect by {\tool}, we also compared the exception types recommended by {\tool} with the one provided by developer. Overall, {\tool} achieves a high level of accuracy in predicting exception types. Over 128 exception bugs, {\tool} recommends adding exception handling blocks for 116 cases (90.62\%). It recommends the exception types in top-1 recommendation for 86 cases (74.13\%), top-2 recommendation for 96 cases (82.75\%), and top-3 for 104 cases (89.65\%).

\subsubsection{Handling Exception Bugs}
In the section, we evaluate the effectiveness of {\tool} in recommending repairing actions for real exception bug fixes. As we focus repairing actions related to Android APIs, we collected a dataset contains 42 exception bug fixes from the same 10 Android open-source projects as in the previous section. In each bug fix of the dataset, developers performed at least one repairing action (i.e. a method call, an assignment, etc.) in the exception handling code. Figure \ref{empirical:example1} in the previous section shows an example of a bug fix in the dataset. In the bug fix, the developer performed three method calls and one assignment to set value for \code{postContent}.

\begin{table}[h]
	\centering
	\sf
	\caption{Result in recommending repairing actions}
	\label{second:table}
	\begin{tabular}{lrr}
		\toprule
		& ExAssist & Barbosa \textit{et al.} \\
		\midrule
		The number of exception bug fixes                  & 42        &    42\\
		The number of matches  & 27        &  6    \\
		The number of partial matches & 3            &  3\\
		The number of misses        & 12          &    33\\
		\bottomrule
	\end{tabular}
\end{table}

For each bug fix in the dataset, we invoked {\tool} to recommend repairing actions and compare the recommendation result with the corresponding fix. 
%Note that, as a recommendation tool, we has improved the functionality of {\tool} by not only recommending repairing method calls but also recommending other actions such as assigning a default value for a variable, returning a default value, etc. 
If the recovery actions recommended by {\tool} exactly match with the fix, we count as a match. If the recommended actions of {\tool} contain the code in the fix of the developer, we count as a partial match. Otherwise, we consider the case as a miss. 

To further evaluate the effectiveness of {\tool}, we compare our tool with a well known existing
approach provided by Barbosa \textit{et al.} \cite{barbosa_heuristic}. In their work, they proposed a technique that uses exception types, method calls, and object types as heuristic strategies to identify and recommend relevant code examples with current editing handling code. We re-implemented the technique using the same configurations presented in the work. 

Table \ref{second:table} shows the evaluation result of {\tool} and the baseline for the task. In the total of 42 exception bug fixes, the recommendations of {\tool} match the fixes of developers in 27 cases. There are 3 cases in which the recommended actions of {\tool} contain the fix code. Overall, {\tool} could provide meaningful recommendations in roughly 64\% of the bug fixes. The result of {\tool} also outperforms the baseline method substantially as the baseline only match the fixes of developers in 6 cases.

\section{Related Work}

There are several studies empirical studies on exception handling \cite{Ebert_exception} and \cite{hazard}. Ebert \textit{et al.} \cite{Ebert_exception} presented an exploratory study on exception handling bugs by surveying of 154 developers and an analysis of 220 exception handling bugs from two Java programs, Eclipse and Tomcat. Coelho \textit{et al.}~\cite{hazard} performed a detailed empirical study on exception-related issues of over 6,000 Java exception stack traces extracted from over 600 open source Android projects.

P\'{a}dua \textit{et al.}~\cite{dePadua_post} investigated the relationship between software quality measured by the probability of having post-release defects with exception flow characteristics and exception handling anti-patterns. In \cite{Padua_revisit}, they studied exception handling practices with exception flow analysis. Kechagia \textit{et al.}~\cite{Kechagia_riddle} investigated the exception handling mechanisms of the Android platforms’ API to understand when and how developers use exceptions. In \cite{Kechagia_toward}, they examined Java exceptions and propose a new exception class hierarchy and compile-time mechanisms that take into account the context in which exceptions can arise. In \cite{Kechagia_undocumented}, they showed that a significant number of crashes could have been caused by insufficient documentation concerning exceptional cases of Android API. Bruntink \textit{et al.}~\cite{Bruntink_idiom} provided
empirical data about the use of an exception handling mechanism based on the return code idiom in an industrial setting. Coelho \textit{et al.}~\cite{Coelho_unveiling} studied exception handling bug hazards in Android based on GitHub and Google code issues. In \cite{Melo_unveiling}, they studied exception handling guidelines adopted by Java developers.

Exception handling recommendation has been studied in several researches~\cite{rahman_context,barbosa_heuristic,barbosa_global,barbosa_enforce,barbosa_awareness,filho_prevent}. Barbosa \textit{et al.}~\cite{barbosa_heuristic} proposed a set of three heuristic strategies used to recommend exception handling code. In \cite{barbosa_global}, they proposed RAVEN, a heuristic strategy aware of the global context of exceptions that produces recommendations of how violations in exception handling may be repaired. Rahman \textit{et al.}~\cite{rahman_context} proposed a context-aware approach that recommends exception handling code examples from a number of GitHub projects. Filho \textit{et al.}~\cite{filho_prevent} proposed ArCatch, an architectural conformance checking solution to deal with the exception handling design erosion. Lie et al.~\cite{Liu_expsol} proposed an approach, named EXPSOL, which recommends online threads as solutions for a newly reported exception-related bug.
%  In \cite{barbosa_enforce}, they presented a DSL to specify and verify exception handling policies.

There exist several methods for mining exception-handling rules \cite{Zhong_repair,weimer_mistakes,thummalapenta_carminer}. WN-miner~\cite{weimer_mining} and CAR-miner~\cite{thummalapenta_carminer} are approaches that use association mining techniques to mine association rules between method calls of \code{try} and \code{catch} blocks in exception handling code. Both models are used to detect bugs related to exceptions. % Zhong \textit{et al.}~\cite{Zhong_repair} proposed an approach named MiMo, that mines repair models for exception-related bugs.

There are several fuzzy-based approaches have been proposed to solve problems in software engineering, such as, bug triaging problem~\cite{fuzzy_tamrawi,fuzzy_riaging}, automatic tagging~\cite{fuzzy_kofahi}, bug categorization~\cite{fuzzy_chawla}. Statistical language models have been successfully used to capture patterns in source code \cite{tu_localness,allamanis_massive,NATURALIZE,jacob_template,maddison_tree,ngram}. For example, Hindle \textit{et al.} \cite{hindle_naturalness} shows that source code is repetitive and predictable like natural language. SLAMC \cite{SLAMC} represents code by semantic tokens.

Several tools for API usage recommendation have been developed but most of them only focus on normal usage of objects. Grapacc~\cite{grapacc} recommends API usage patterns with frequent graph-based models. DroidAssist~\cite{nguyen_droidassist} recommends on method calls based on Hidden Markov Models.

\section{Conclusion}
Exceptions are unexpected errors occurring while an app is running. Learning to handle exceptions correctly is often challenging for mobile app developers due to the fast-changing nature of API frameworks for mobile systems and the insufficiency of API documentation. We propose two techniques for learning and recommending exception types and repairing method calls for exception handling code. Our evaluation shows that our techniques can effectively learn exception handling patterns from a large repository of mobile apps and provide recommendations with a high level of accuracy. Based on the proposed techniques, we have developed {\tool}, a code recommendation tool for exception handling.

\bibliographystyle{IEEEtran}
\bibliography{IEEEabrv,exception17}

\end{document}